# Semiconducting Black Phosphorus: Synthesis, Transport Properties and Electronic Applications

Han Liu, Yuchen Du, Yexin Deng and Peide D. Ye

School of Electrical and Computer Engineering and Birck Nanotechnology Center,

Purdue University, West Lafayette, IN 47907, USA

Email: yep@purdue.edu

**Abstract**

**Phosphorus is one of the most abundant elements preserved in earth, constructing with a fraction of ~0.1% of the earth crust. In general, phosphorus has several allotropes. The two most commonly seen allotropes, white and red phosphorus, are widely used in explosives and safety matches. In addition, black phosphorus, though rarely mentioned, is a layered semiconductor and have great potentials in optical and electronic applications. Remarkably, this layered material can be reduced to one single atomic layer in the vertical direction owing to the van der Waals structure, known as phosphorene, where the physical properties can be tremendously different from its bulk counterpart. In this review article, we trace back to the 100 years research history on black phosphorus from the synthesis to material properties, and extend the topic from black phosphorus to phosphorene. The physical and transport properties are highlighted, aiming at further applications in electronic and optoelectronics devices.**

# 1. Introduction

In the past several years, we have witnessed the great progress in the research field of semiconducting 2D crystals. Most of the 2D semiconductors lie in the family of a series of transition metal dichalcogenides (TMD) in the form of "$MX_2$", where M represents a transition metal such as Mo or W, and X can be either S, Se or Te. These 2D semiconductors have been changing the landscapes of many research fields in science and technologies, particularly in condensed matter physics, electronic devices, chemical sensors and energy storages. [1-9] Tremendous efforts are still ongoing to uncover the full potential of TMDs and their derivatives.

The most distinctive property of 2D semiconductors arise from their unique van der Waals structure. The bulk crystals are consisted of individual layers, where each layers are vertically stacked by the van der Waals force instead of covalent bonds. This leads to the possibility in reduction of the crystal thickness in the direction perpendicular to the 2D plane. As a result, the crystal thickness can be scaled down to the atomic layer scale with significant changes in the physical properties. For example, a transition from indirect to direct band gap in $MoS_2$ is observed from bulk to single layer with a larger band gap. [1,2] Also, this atomic-layer thin channel is much favored for the aggressively scaled field-effect transistors avoiding so called short channel effect, an undesirable phenomenon when the dimension of the transistors is aggressively scaled. [6,7]

In addition to TMDs, it is interesting to explore new 2D semiconductors. Black phosphorus, a rare allotrope of phosphorus, is one of the promising candidates. The discovery of black phosphorus can be dated back a century long ago. It was first synthesized from red phosphorus under high temperature and pressure. [10] In the past century, the synthesis, physical properties, and device applications have been extensively investigated in various studies. Herein, we take a retrospective review to the discussion of fundamental properties of this material. Remarkably, the very recent successful demonstration of single layer black phosphorus, known as "phosphorene", has rendered renaissance to this material. [11-13] In this review, we aim to cover studies from early stages to present on black phosphorus, including synthesis and carrier transport properties. We also underscore the recent development of material properties of few to single layer phosphorene and provide an insightful comparison between the bulk and few layer materials from fundamental properties to carrier transport, in order to explore the full potential of electronic and optoelectronic device applications of this 2D semiconductor.

## 2. Synthesis and Crystal Structures

*2.1 SYNTHESIS*

Black phosphorus was first synthesized early in 1914 by applying high hydrostatic pressure at a temperature below which the transformation runs with appreciable velocity at atmospheric pressure. [10] White phosphorus was placed in a high pressure cylinder under kerosene. High pressure up to 0.6 GPa was applied at room temperature. The cylinder was then elevated to 200 ℃ and pressure then up to 1.2 GPa. It took 5-30 min for the transition from white phosphorus to black phosphorus occurred under this circumstance. A small volume of black phosphorus was found after cooling the cylinder and relieving the pressure. As the most striking difference, Bridgeman noticed that black phosphorus showed high density of 2.69 g/cm$^3$ compared to 1.83 g/cm$^3$ of white phosphorus and 2.05-2.34 g/cm$^3$ of red phosphorus. Also, unlike white or red phosphorus, it showed its chemical stability as the incapability to catch fire in air or be ignited by fire, and could sustain up to 400 ℃ in air without spontaneous ignition. [14]

Same result was reproduced by Keyes under the hydrostatic pressure of 1.3 GPa at 200 ℃. The ingot achieved from this method was polycrystalline and the grain size was roughly around 0.1 mm. [15] It was also demonstrated that black phosphorus was a good conductor of both electrons and heat but less diamagnetic than white or red phosphorus. The solid structure of black phosphorus by the Bridgman method has been known to change from amorphous to polycrystalline form depending on the applied pressure and temperature. [16] Further, black phosphorus was prepared by using mercury or bismuth-flux method. [17-19] Brown and Rundqvist revealed that it crystallizes in orthorhombic form and has an infinite puckered layer structure. White phosphorus was purified with 15% $HNO_3$ and then solved in liquid bismuth. The mixture was heated up to 400 ℃ and kept for 20 hours and slowly cooled down to room temperature. By dissolving liquid bismuth with 30% $HNO_3$, needle or rod like black phosphorus can be achieved. The size of a typical crystal was of 5×0.1×0.07 mm$^3$ with needles along the a-axis. Silicon or germanium doped black phosphorus can also be achieved by bismuth-flux method. [20,21] In 1981, large single crystals of black phosphorus was first succeeded from red phosphorus melted at high temperature under high pressure using a wedge-type cubic high-pressure apparatus. Samples grown under 3.8 GPa higher than 270 ℃ were crystalized in a single crystal. [22]

It was not until very recently that black phosphorus was reported to be achieved at lower pressures. Lange and colleagues reported a surprisingly low-pressure route to produce high-quality black phosphorus by using a mineralizer as reaction promoter at non-toxic conditions. Black phosphorus was prepared by the reaction of Au, Sn, red phosphorus, and SnI$_4$ in evacuated silica ampules. The starting materials were heated to 600 ℃ and gradually cooled down to room temperature. This reaction usually took several dozen of hours up to 5-10 days to complete. [23-24] This process is further simplified by Köpf using Sn/SnI$_4$ as the only mineralization additive. [25]

*2.2 CRYSTAL STRUCTURE*

At normal conditions, bulk black phosphorus is a layered structure and assembles graphite. The crystal structure is shown in Figure 1. The unit cell contains eight atoms, which gives a calculated density of 2.69 g/cm$^3$. With black phosphorus prepared by Bridgman method, the lattice constant was first determined by Hultgren *el al.* with X-ray pattern. This unit cell contains two layers. Each phosphorus atom is bonded with three neighboring atoms at 2.18 Å. Two of them are in the plane of the layer at 99° from one another, and the third is between the layers at 103°, making an average angle of 102°. [26] Similar results were reproduced by Brown and Rundqvist with black phosphorus crystals from bismuth-flux method. [18] STM studies reveal that the atoms occupy almost the same position on the surface, with a little relaxation perpendicular to the surface, which causes the atoms to take two types of positions with a contrast visible in the STM images. A small difference in height of 0.02 Å determined by DFT calculations breaks the symmetry, which gives difference to the band structure. [27] Lattice parameter extractions at normal conditions were done by Brown and Rundqvist and are listed in Table 1. [18]

Table 1: Crystal structural parameters of black phosphorus under the normal condition

| | | |
|---|---|---|
| a | 4.374 Å | lattice constants in orthorhombic system |
| b | 3.3133 Å | |
| c | 10.473 Å | |
| u | 0.0806 Å | crystal structural parameters |
| v | 0.1034 Å | |
| $d_1$ | 2.222 Å | bond length |
| $d_2$ | 2.277 Å | |
| $\alpha_1$ | 96.5 ° | bond angle between $d_1$s |
| $\alpha_2$ | 102.09 ° | bond angle between $d_1$ and $d_2$ |

The crystal structure of black phosphorus can be greatly discriminated under high pressure. [28] This mostly comes from the anisotropic compressibility of black phosphorus, due to the asymmetrical crystal structures. It can be expected that in the z-direction, the van der Waals bond can be greatly compressed. However, it also shows a strong variation in compressibility across the orthogonal x-y plane. The a-axis shows similar compressibility at relatively lower pressure (<2.66 GPa) with c-axis, where along the b-axis, it almost remains constant. This essentially results in structural change for black phosphorus under high pressure. [29,30] As a matter of fact, black phosphorus undergoes two reversible structural transitions at high pressures. The first transition, from orthorhombic to rhombohedral phase occurs around 5.5 GPa at room temperature, accompanied by displacement of the puckered layers and a volume change. With higher pressure, it goes from rhombohedral to simple cubic phase at around 10 GPa. Transition temperature is almost independent with temperature. [31-32] This simple cubic phase keeps stable even the pressure is increased up to 60 GPa. [33]

## 3. Carrier Transport in Black Phosphorus

*3.1 ELECTRICAL CONDUCTION AND CARRIER MOBILITY*

The electronic structure of black phosphorus had been determined by various approaches, including the tight-bonding method, self-consistent pseudopotential method and local-orbital method. [34-36] From band structure calculation, it can be seen that for both the electron and the hole, the effective mass in the x-y plane is fairly large. The effective mass in the x-direction is lightest, and that in the layer-stacking z direction is lighter than that in the in-plane y-direction due to the van der Waals structure. An average effective mass of 0.22 $m_0$ for electrons and 0.24 $m_0$ for holes determined by $(m_x m_y m_z)^{1/3}$ were determined by self-consistent pseudopotential method. Both were slightly smaller than those in silicon. [35]

Electrical conduction of black phosphorus was first measured by Bridgman. His results showed the resistivity roughly ranges between 0.48 and 0.77Ω·cm at 30 ℃, and to be about 35% less at 75 ℃. [37] It was later revisited by Keyes and Warchauer using polycrystalline crystals prepared by Bridgman method. [15,38] Both had determined that undoped samples are showing p-type conductivity with positive Hall coefficient. This indicated that conductivity is more dominated by holes instead of the electrons. Temperature dependent conductance study revealed an energy gap of around 0.33 to 0.35 eV. Later in the 1980s, both n-type and p-type black phosphorus were successfully synthesized and studied. The n-type black phosphorus was realized by Te-doping. N-type conductivity was attributed to the substitution of Te atoms to the P atoms in black phosphorus, however, p-type conductivity is so far not clear. The effective donor and acceptor concentration in these samples were determined to be around 2 to $3 \times 10^{16}$ and 2 to $5 \times 10^{15}$ cm$^{-3}$, respectively. The activation energy for n- and p-type black phosphorus was determined to be ~39 and ~18 meV. [39] Baba *et al.* observed two types of acceptors, of which activation energies were 26.1 meV and 11.8 meV in bismuth-flux prepared samples. Effective concentrations of acceptors were typically $1.36 \times 10^{15}$ cm$^{-3}$ for the deeper level and $0.44 \times 10^{15}$ cm$^{-3}$ for the shallower level. [40]

At room temperature, hole and electron mobility are of 350 cm$^2$/V·s and 220 cm$^2$/V·s respectively. They can be characterized with the T$^{-2/3}$ relationship, indicating that they are mostly scattered by lattice vibration in the measurement range of 195 ℃ to 350 ℃. [15] The resistivity was observed to increase near liquid-helium temperature, and the Hall coefficient peak occurred between 24 and 30 K. [38] Maruyama *et al.* measured the resistance between 1.4 and 400 K and it turned out to be an impurity dominated semiconductor. A band gap of 0.31 eV was estimated. Carrier mobilities extracted from the magnetoresistance coefficients gave the values of $2.7\times10^4$ cm$^2$/V·s at 77 K and $1.5\times10^4$ cm$^2$/V·s at 294 K, respectively, which are roughly one order of magnitude larger than those found from the Hall effect on polycrystalline samples. At 20 K, Hall mobility of holes reached the maximum value of $6.5\times10^4$ cm$^2$/V·s, while the peak shifted to 50 K for electrons around $1.5\times10^4$ cm$^2$/V·s. [41] Magnetoresistance was also studied in these samples down to 0.5 K in temperature and up to 6 T in magnetic fields, and 2D Anderson localization was also observed. [42] In addition, the electric conductivity showed a log T-like behavior below about 5 K and the Hall coefficient had its sign reversal at around 7 K. Magneto-transport properties observed at low temperatures in black phosphorus crystals is originated from the 2D Anderson localization in an inversion layer on the surface. [43]

*3.2 ELECTRICAL CONDUCTION UNDER HIGH PRESSURE*

Bridgeman first noticed that the electrical conductivity of black phosphorus was very sensitive to applied pressure. Under 1.2 GPa, the resistivity is only ~3% of its value under atmospheric pressure. [37,44] Pressure effect on electrical conductivity was further investigated with more details by Okajima *et al.* with single crystal black phosphorus. [45] Akahama et al. measured the resistivity at room temperature in terms of the applied pressure, which decreases logarithmically at higher pressure. Two anomalies were observed, occurring at 1.7 and 4.2 GPa. The first one is mostly attributed to the change of the band gap, and the second one is attributed to the phase change from orthorhombic to rhombohedral. The energy gap decreases at higher pressure and finally became zero around 1.7 GPa, and resistivity turned out to be the characteristic feature of a metal, showing a pressure-induced semiconductor-metal transition. [39] Calculation of energy gap under pressure based on self-consistent pseudopotential method indicated a change of -212 to -235 meV/GPa. [35,46] The linear change of band gap under pressure was attributed to overlap between conduction and valence bands without any structural change.

*3.3 SUPERCONDUCTIVITY*

As mentioned above, black phosphorus shows an interesting sequence of structural transformation. At about 5 GPa, it changes from semiconducting orthorhombic phase to semiconducting rhombohedral, and to metallic simple cubic phase at about 10 GPa. Superconductivity was first observed by Wittig *et al.* near 4.7 K at the simple cubic phase. [47] This constituted one of the four last missing links in the proof that superconductivity is normal behavior for every truly metallic sp element besides arsenic, sulfur, and iodine. The following three paths was studied on dependence of treatments of cooling and pressurizing were studied. [46] (1) First apply 15 GPa at room temperature to convert the samples to simple cubic phase, and then the pressure cell was cooled down to liquid helium temperature. The pressure dependent superconducting transition temperature $T_c$ was then measured. $T_c$ was measured to be around 6 K and increased slightly with increasing pressure. Even after removing the applied pressure it remains superconducting at 6 K for several hours. (2) Pressure was first applied at 8.7 GPa at room temperature where samples are transformed into rhombohedral phase. Pressure cell was then cooled down to liquid helium temperature and then pressure dependent $T_c$ was measured. Transition observed at 8.7 GPa suggested the possibility of superconductivity in the rhombohedral phase. With increasing pressure, transition curve at about 25 GPa resembles previous path, suggesting the transformation from rhombohedral to simple cubic may occur even at liquid helium temperature. (3) Samples were first cooled down to liquid He at normal pressure. Pressure dependent $T_c$ was measured by applying high pressures. $T_c$ rose from 4 K to 10.7 K when pressure increased from 11 to 30 GPa. Also, the transition temperature decreases rapidly by decreasing the applied pressure and becomes null before about 3 GPa in contrast with the first path. [47,48]

Two mechanisms were applied to explain the superconductivity in black phosphorus. One is related with dislocations presented in black phosphorus crystals. Atoms around these dislocations are relatively more mobile, thus transition from the orthorhombic to the simple cubic phase via the rhombohedral phase would take place around dislocations even at low temperature. Therefore, black phosphorus is probably converted to semiconducting phase matrix with metallic fine threads as a mixed system. The appearance of superconductivity in a system like this might be possible. [49] Second, as mentioned above, the energy gap is relevant with applied pressure. The band gap vanishes at about 1.7 GPa with a metallic state transition without structural change. The superconductivity observed in the third path of black phosphorus is unique, which shows that the material property can be significantly changed under high pressure. [46]

## 4. Device Applications

*4.1 FIELD-EFFECT TRANSISTORS*

100 years after its discovery, black phosphorus research gets its renaissance following the trend of researches on device applications with TMD channel materials. Field-effect transistors have also been demonstrated with black phosphorus channels. [11-13,50,51] The thickness of the black phosphorus crystal used as the channel material ranges from 2-30 nm. Figure 3(a) shows a schematic image of a typical back-gated black phosphorus transistor. Heavily doped silicon was used as the back gate with 90 nm $SiO_2$ on top of that as the back gate dielectric. Few nanometer thick black phosphorus crystals were cleaved from bulk ingots are transferred on the substrate. Source/drain contacts were lithographically defined and followed by metallization and lift-off. Output and transfer characteristics are shown in Figure 3(b) and (c). A maximum drain current of 194 mA/mm was achieved at 2 V drain bias for the 1 μm channel length transistor. This is a significant improvement of drain current for transistors based on 2D semiconductor channel materials, where the drain current is almost an order of magnitude larger than the most reported TMD channel transistors at a similar gate length. A high field-effect mobility from several hundred up to 1000 cm$^2$/V·s at room temperature is estimated from maximum transconductance.[12] Both field-effect mobility and Hall mobility show a temperature dependence, as shown in Figure 4. They decrease at temperatures higher than ~100 K, and saturate at lower temperatures. It is attributed to electron-phonon scattering between the range from 100 to 300 K and can be characterized by the $T^{-1/2}$ feature, and impurity scattering dominates in the range lower than 100 K. One thing interesting to mention is, despite a prominent p-type conductivity described in the previous section, n-type conductivity can also be observed by the field effect. [12,52] In consistency of positive Hall coefficient at negative gate bias, a sign reversal of Hall coefficient was observed at high positive electrical field (>6.6 MV/cm). This is attributed to the narrow band gap nature in black phosphorus, where the Fermi-level can be tuned either near the valence band, leading to the hole conduction, or the conduction band, leading to the electron conduction.

Detailed device perspectives were also studied by Du *et al.*, where contact resistance and scaling properties were discussed in the great details. [53] Contact resistance of Ni and Pd on black phosphorus saturates at around 1.75±0.06 and 3.15±0.15 Ω·mm at electrostatic doping limit. Notably, as shown in Figure 5(a) of typical transfer curves with various channel lengths, n-type conductivity is enhanced at shorter channel lengths. This enhancement of n-type conductivity is attributed to the Schottky barrier nature in transistors based on 2D semiconductor channels. [54] Away from silicon MOSFETs, carrier transport in TMD based transistors are greatly influenced by the Schottky barriers at the source/drain contacts. This prevails in black phosphorus transistors as well. As Fermi-level of the metal contact is aligned at the vicinity of the valence band in black phosphorus, p-type conduction is easier to be observed in black phosphorus transistors, due to a smaller Schottky barrier for holes than electrons. As shown in Figure 5(b) and (c), considering band bending, the effective barrier height is only determined by gate bias for long channel transistors. Once channel length is reduced, the effective barrier height can be modulated by drain bias as well. Smaller effective barrier height realized by larger drain bias would facilitate electron injection from metal source to channel semiconductors. Slightly different from the effect of drain induced barrier lowering (DIBL) in short channel silicon MOSFETs, in black phosphorus transistors, it alternates the effective barrier height at the "on-states" of the transistors instead of "off-states", and leads to an enhanced on-state drain current.

Admittedly, black phosphorus enjoys several advantages from the devices aspects. First of all, black phosphorus is mainly p-type semiconductor which provides a good addition to 2D family since most of TMDs are n-type. It's easy to constitute the 2D CMOS [11], the 2D p-n junction [55] and the 2D heterojunction tunneling field-effect transistor. Secondly, it has relatively higher carrier mobility compared to TMDs. Thirdly, its unique anisotropic transport properties might provide fundamental material advantage as channel in device applications at ultimately scaled ballistic limit, which we will discuss below. [56] However, the further enhancement of device performance is limited by the following bottlenecks, which need to be seriously addressed in the near future.

The first problem comes from the optimization of the crystal thickness. It has been found that thickness of the crystal plays a critical role in device performance, and both current on/off ratio and field effect mobility are extremely sensitive to crystal thickness. [11,13] Thicker crystals would yield high driving current, but in the meantime significantly decreases the transistor current on/off ratio. This gives challenge to the determination of the crystal thickness considering the applications of the transistors, and also gives challenge to further large scale synthesis of black phosphorus, as even tiny deviation in crystal thickness uniformity would result in large variation of device performance.

The other issue comes from the surface instability of black phosphorus. This has been observed in several studies regarding the isolation of black phosphorus thin layers. [50,57] The mechanism of surface degradation is not well understood yet. It is so far considered that pure black phosphorus is a chemically stable material, but the existence of tiny traces red phosphorus from incomplete reaction during synthesis of black phosphorus would induce the surface oxidation. Favron *et al.* used Raman spectroscopy as a probe of the quality and integrity of exfoliated layers to present a dynamical study of the photo-oxidation process at room temperature in a controlled atmosphere with the presence of the oxygen-water redox couple. A photo-induced charge transfer from black phosphorus to the redox couple is found to be responsible of the fast deterioration of the structure. [58] Later DFT study by Ziletti *et al.* shows oxygen defects in black phosphorus which not only serve as low energy metastable forms and introduce deep donor and/or acceptor levels in the gap, but also increase the hydrophilicity of phosphorene due to their ability to establish hydrogen bonds, and hence degrade the black phosphorus surface. [59]

*4.2 ANISOTROPIC TRANSPORT*

The anisotropic transport was recorded in many early studies of black phosphorus. [22,39,60] The z-direction perpendicular to the x-y plane was showing the least carrier mobility, and the x-direction in the 2D plane perpendicular to the ridges was showing the highest carrier mobility. It is interesting to note that z-direction has the lowest mobility, but not the highest effective mass. It is not strictly following the expectation from its effective mass. For electrons, $m_x=0.0826m_0$, $m_y=1.027m_0$, $m_z=0.128m_0$ and for holes, $m_x=0.076m_0$, $m_y=0.648m_0$, $m_z=0.28m_0$ were expected. [61] However, the smallest mobility occurs in the z-direction, while the largest in the x-direction. This might be attributed to the fact that inter-layer transport is significantly different from intra-layer transport. Different results of $m_x=0.3m_0$ for electron/hole and $m_y=2.6m_0$ for electrons, $8.3m_0$ for holes are obtained by recent DFT calculations. [11] This anisotropy is also confirmed in field-effect black phosphorus transistors. Both drain current and field-effect mobility were characterized by the transport direction. Despite the limited 45° angular resolution, the observed 50% anisotropy is still significant, and has not be observed in any bulk semiconductors. [11] This indicates that the crystal orientation can be identified not only by high resolution electron microscope, but also by electrical transport.

In addition to the measurement of electrical conductivity, the orientation of the crystals can also be identified by angle-resolved optical extinction spectrum measurement. [13] As shown in Figure 6, extinction spectra with the incident light from the z-direction at six linear polarization angles separated by 30 degrees was measured on a black phosphorus crystal. It showed a sharp increase at around 2400 cm$^{-1}$, indicating a band gap of around 0.3 eV, in good agreement with black phosphorus band gap at Z symmetry point. Since the optical conductivity of black phosphorus peaks at the x-direction at the band edge, the anisotropy in optical conductivity arises from the directional dependence of the interband transition strength in anisotropic black phosphorus bands. Corresponding electrical measurement further determined a ratio of mobility along x- and y-directions $\mu_x/\mu_y$ of 1.5, which is similar to the ratio between maximum and minimum values in angular dependent conductance measurement. Moreover, direction oriented photocurrent was also observed. [62] The device advantage using this unique anisotropic transport needs to be further demonstrated. In principle, we can design the transport along x-direction which provides highest mobility and channel width along y-direction which has the highest density of states where density of states remains the same determined by $(m_x m_y)^{1/2}$ [56] It is superior over other conventional III-V semiconductors which has high electron mobility but low density of states. We can also design unique electro-optical and electro-thermal devices using this anisotropic transport property. [63,64]

*4.3 OPTOELECTRONIC DEVICES*

Black phosphorus is very promising in optical and optoelectronic applications. [65,66] As a direct and narrow band gap semiconductor, p-type black phosphorus enjoys plenty of advantages in future applications as building blocks for functional optoelectronic devices. Different from $MoS_2$ with a ~1.8 eV direct bandgap only in its monolayer form, black phosphorus shows a thickness-dependence direct bandgap, which exhibits ~0.3 eV in its few-layers form and increases to >1 eV in its single layer form. Its relatively small direct band gap makes it ideal for infrared optoelectronic application. Combined with its high mobility, it shows promise for fast and broadband photodetection and solar cell applications. Multi-layer black phosphorus photo-detector has been realized by Engel *et al.* [67]. Multi-spectral, high-resolution imaging has been demonstrated in both visible ($\lambda_{VIS}$=532nm) and near-infrared ($\lambda_{IR}$=1550nm) regions.

P-N diode is the basic element of both electronic and optoelectronic devices. Black phosphorus p-n diode has been demonstrated based on homojunction black phosphorus or the combination of p-type black phosphorus and n-type $MoS_2$ van der Waals heterojunction recently, as shown in the inset of Figure 7(a). [55,68] Black phosphorus was mechanically exfoliated onto CVD monolayer $MoS_2$ to form the van der Waals heterojunction. The highly doped Si substrate and $SiO_2$ were used as back gate and gate dielectric, which can modulate the p-n junction electrostatically. The p-n diode shows a strong gate tunable current-rectifying characteristics, indicating a good p-n junction was formed. When used as a photodiode, it shows a strong photoresponse to a 633 nm incident laser, and the photocurrent increases as the incident laser power increases, which can modulated by the back gate, as indicated by Figure 7(a) and (b). Photoresponsivity (R) is used to describe the sensitivity of the photoresponse of the device, which is defined as $R=I_{ph}/P_{laser}$ where $I_{ph}$ is the photocurrent and $P_{laser}$ is the power of incident light. The maximum R in this p-n diode is up to 3.54 A/W (at $V_d$ = +2 V) and 418 mA/W (at $V_d$ = -2 V) under 1 µW laser power. As a photodiode, this is nearly 100 times higher than that of the recently reported black phosphorus phototransistor, and 4.8 times higher than that of the carbon nanotube-$MoS_2$ p-n diode with a much smaller $V_d$. [69,70] By using this electrically tunable 2D p-n diode, it is possible to realize sensitive and broadband photodetection due to the high mobility and relatively small direct band gap of few-layer black phosphorus. Moreover, this p-n diode can also be used as solar cell. As shown in Figure 7(c) and (d), under laser illumination, the short circuit current ($I_{sc}$), and the open circuit voltage ($V_{oc}$) increase as the laser power increases. The power generated by the p-n diode can be calculated by $P_d = I_d V_d$. Then, the fill factor, which is defined as FF=$P_d/(I_{sc}V_{oc})$, can be obtained. The maximum FF is ~0.5 in this 2D p-n diode. The external quantum efficiency (EQE) can be calculated by EQE = $(I_{sc}/P_{laser})$ $(hc/e\lambda)$, where h is the Planck constant, e is the electron charge, c is the velocity of light, $\lambda$ is the wavelength of the incident light. The peak EQE in this device is 0.3%. This is the first time that efficient photovoltaic energy conversion has been demonstrated using black phosphorus. By further reducing the thickness of the black phosphorus to bilayer, it is possible to essentially enhance the EQE to up to 18%, according to theoretical predictions. [71] Moreover, by stacking the 2D p-n diode vertically, a stacked solar cell structure can be realized. Since the direct band gap of black phosphorus changes with its thicknesses, the efficiency may be further improved by tuning its thickness to better utilize the photon energies of different wavelengths of the solar spectrum. Photocurrent mapping was performed on this device, which found out that the photocurrent is generated across the whole overlapped region. This make this p-n diode possible for large area application if transparent electrode (e.g. graphene) can be used.

Phototransistor was also demonstrated on black phosphorus. The first phototransistor was demonstrated with an R of 4.8 mA/W and a response time of 1 ms. [69] It shows a response to visible light up to 940 nm. However, the relatively low photoresponsivity calls for more attention on device engineering. Metal contacts on 2D materials is one of the significant issues of improving 2D device performances.

Based on the experiment results on black phosphorus field-effect transistor with various metal contacts with different work functions, it shows that the work function of metal plays an important role on the hole/electron conductions at the contact. Field-effect transistor with larger work functions metals as contacts exhibit larger hole current, while ambipolar characteristics can be observed on devices with lower work function metals. In order to reduce the Schottky barrier height at the metal contact/black phosphorus, Pd (5.1 eV) is used as metal contacts. The maximum R in these device is 223 mA/V at $V_{gs}$=-30 V and 76 mA/W at $V_{gs}$=0. This value is 16 times larger than the value reported in the first black phosphorus phototransistor at $V_{gs}$=0 and with an even smaller $V_{ds}$.[55]

## 5. Phosphorene: from Bulk to Single Layer

Plenty works regarding fundamental properties and device applications have been done from bulk to few-layer black phosphorus and reported since reference 11 and 12 were posted on arXiv in January 2014. [72] On seeing the triumph from graphene whose properties completely distinguishes from bulk graphite, the next question is, is it possible to isolate single layer black phosphorus, known as phosphorene, and how would its properties distinguish from bulk crystals? In addition, we are still wondering if there would exhibit new physics, new device applications, and surprises.

Isolation of single layer phosphorene has already been demonstrated from classical mechanical exfoliation or plasma etching. [11,73] Photoluminescence and Raman spectrum were first performed to unveil the difference in physical properties. Larger band gap in few-layer phosphorene has been observed with reduced layer numbers by photoluminescence. [74] A pronounced peak centered at around 1.45 eV was observed in single layer phosphorene, which indicates that band gap structure in few- or single-layer phosphorene differs from bulk crystals. Moreover, obvious Raman peak shifts in single- and bi-layer phosphorene and bulk black phosphorus were observed as well. [11]

The bulk black phosphorus is a narrow band gap material. Beside optical measurement, recent studies have confirmed a 0.2 to 0.3 eV direct band gap by Angle-resolved Photoemission Spectroscopy. [12,75] However, due to the surface oxidation, as discussed in previous sections, the preparation of single layer phosphorene sample for ARPES might undergo some difficulty at this moment, making a precise determination of single layer phosphorene band structure not available at this moment. Nevertheless, numerical methods would help to predict the unique properties of single layer phosphorene theoretically. Single layer phosphorene has been expected to have a much larger band gap than bulk materials, which is similar to that in $MoS_2$. [1] Due to the difference in simulation approaches, the expectation of the band gap of few layer phosphorene slightly increases from bulk values, and the band gap for single layer phosphorene varies between 0.75 eV to 2 eV. [11,76-79] It was also noticed a new phase of black phosphorus, blue phosphorus, has a fundamental large band gap over 1.1 eV for bulk, and this can be even increased to 2 eV for single layer. [80,81]

The band structure is sensitive to strain. An increasing compressive strain would lead to the band gap transiting from direct to indirect (0.6 GPa), and then to zero (1.7 GPa), while the tensile strain would enlarge the band gap almost linearly. In addition, a compressive strain of 1~2% would enhance the thermoelectric performance of electron (n-type) doped black phosphorus, while tensile strain benefits hole (p-type) doped black phosphorus. [64,82] The band gap in single layer phosphorene nanoribbons was also studied. Unlike graphene, both hydrogen terminated phosphorene nanoribbons, either zigzag or armchair nanoribbons, have been predicted to be semiconductors, inheriting the semiconducting properties of phosphorene. [83,84] The zigzag phosphorene has been found to have the greatest quantum size effect, tuning the band gap from 1.4 to 2.6 eV when the width is reduced from 26 to 6 Å. [84] Effect from external electrical field was examined. The theoretical result revealed that, due to giant Stark effect, a transverse electrical field would induce a metal-insulator transition and can eventually close the band gap at high field. The critical value to close the gap depends on the width of the nanoribbon. In sharp contrast, perpendicular electric field failed to change the electronic structures in phosphorene nanoribbons. [83] The band structure of phosphorene nanoribbons can also be changed by chemical decorations on the edge. However, armchair phosphorene nanoribbons remains semiconducting regardless of edge groups, while zigzag phosphorene nanoribbons can shift from semiconducting to metallic depending on edge chemical groups. These edge states induced by edge chemicals are at the bottom of the conduction band and result in a reduced band gap. [85] Meanwhile, the effective mass in phosphorene nanoribbons was tunable due to external strains. A sudden change for electron effective mass at 8% strain for armchair phosphorene nanoribbon or hole effective mass at 3% strain for zigzag phosphorene nanoribbon was determined, leading to a drastic change in $m_e^*/m_h^*$ ratio by an order of magnitude in each case. [84]

Remarkably, single layer phosphorene can be even tuned superconducting with electron doping. [86] When electron doping concentration is above $1.3 \times 10^{14}$ cm$^{-2}$, superconductivity starts showing up in single layer phosphorene. Higher doping concentration elevates the $T_c$ which would saturate at 11.2 K once doping concentration reaches $7.2 \times 10^{14}$ cm$^{-2}$. This is comparable to the simple cubic phase of black phosphorus under high pressure, and might be a new superconducting phase in black phosphors.

## 5. Conclusion and Perspectives

From material aspects, black phosphorus is unique. Unlike other phosphorus allotropes such as red or white phosphorus, black phosphorus is a thermodynamically stable semiconductor. It comes from white or red phosphorus at high pressure and high temperature from a phase transition. Under atmosphere pressure, black phosphorus is an orthorhombic narrow band gap semiconductor. However, this material undergoes phase change at higher pressure to become rhombohedral at around 5 GPa and simple cubic at 10 GPa, accompanied by a semiconductor-metal transition and ultimately becomes a superconductor. The band gap in black phosphorus starts to shrink at higher pressure and finally diminishes at 1.7 GPa. This pressure induced structural change in 2D materials has only been observed in black phosphorus so far.

The unique properties in black phosphorus, along with its high carrier mobility, would make it as a promising material in electronic applications as material for functional devices. People have been making endeavors to create a relatively larger band gap in graphene but to no avails. The advent of semiconducting black phosphorus, has bridged this gap. The band gap in black phosphorus is 0.3 eV, and this gap can be expanded to 1.0 to 2.0 eV, depending on the layer numbers. Field-effect transistors have been demonstrated with few-layer to bulk black phosphorus, and they have shown improved performance over most of TMD based transistors in terms of on-current and carrier mobility. In addition, as a direct band gap material, the range of band gap corresponds to an adsorption spectrum between visible light to infrared. Therefore, the optoelectronic application with black phosphorus is even beyond the capability of single layer TMD materials.

The first synthesis of black phosphorus can be traced back to 1914. After exactly a century, the advent of single layer phosphorene has given renaissance to this old material. For development of material research on black phosphorus in the future, there would probably two immediate directions require further studies.

The first is the isolation of single layer phosphorene, effective passivation and transport studies. There has been a significant difference between the bulk graphite and single layer graphene in terms of the electronic structure and carrier transport behavior. Similar contrast might also occur in the black phosphorus and phosphorene case as well. However, this is limited by the surface degradation in black phosphorus. The surface degradation is mostly induced by air exposure, which is inevitable during device fabrication. The chemical reaction mechanism has been thoroughly studied very recently. [87-89] Oxygen and water are identified as the main source for the degradation. The research focus becomes how to protect, passivate, encapsulate the black phosphorus and phosphorene films for the device fabrication. The ultimate solution is to improve the black phosphorus synthesis to eliminate defects due to the unreacted red phosphorus residues, which has been suspected to cause surface oxidation. An alternative approach is to find a proper way, such as atomic layer depostision (ALD), to passivate the surface of phosphorene to block air exposure or to reduce surface chemical activity. [52,88] Detailed temporal and thermal stability of ALD $Al_2O_3$-passivated few-layer phosphorene transistors shows the drain current was stable between -50 $^o$C and 150 $^o$C up to at least 1000 h. [88, 90] With the appropriate passivation process, we can eventually investigate further the transport properties in single layer phosphorene, probe new physics if any, and to discover more potential applications within this fascinating material.

The second direction worth investigating is the hetero-integration of various 2D crystals from graphene, boron nitride, TMDs, and phosphorene. Benefited from their 2D nature, isolated atomic layers can also be reassembled into van der Waals heterostructures made in a precisely designed sequence. [91] Compared to conventional heterojuctions, usually realized by molecular beam epitaxy, the advantage of 2D heterojuction includes: 1) the vertical thickness of each layer can be controlled at atomic scale with perfect interface, where molecular beam epitaxy (MBE) inevitable requires plenty of buffer layers to release strains from lattice mismatch, and 2) material system of heterojunctions stacked by atomically thin crystals would facilitate the study on interlayer coupling, and hence to build up more powerful devices such as lasers, light-emitting diodes, solar cells, and high-electron-mobility transistors. [92] P-N junctions, field-effect transistors and bipolar transistors have been demonstrated so far with these 2D materials, and there will be plenty of new device and designs ahead.

The research on 2D materials, from graphene, TMD to black phosphorus will continue to thrive due to their unique crystal structures from symmetry breaking. This helps to investigate new device physics in these materials and generates new device concepts, and would definitely facilitate more powerful functional devices in the future. This article aims to take a retrospect review from synthesis methods, fundamental properties to novel findings of black phosphorus. We hope this will inspire more exciting discoveries and applications in this growing family of van der Waals materials.

This work is partly supported by NSF under Grants CMMI-1120577, ECCS-1449270 and ARO W911NF-14-1-0572 monitored by Dr. Joe X. Qiu.

**Figure 1**

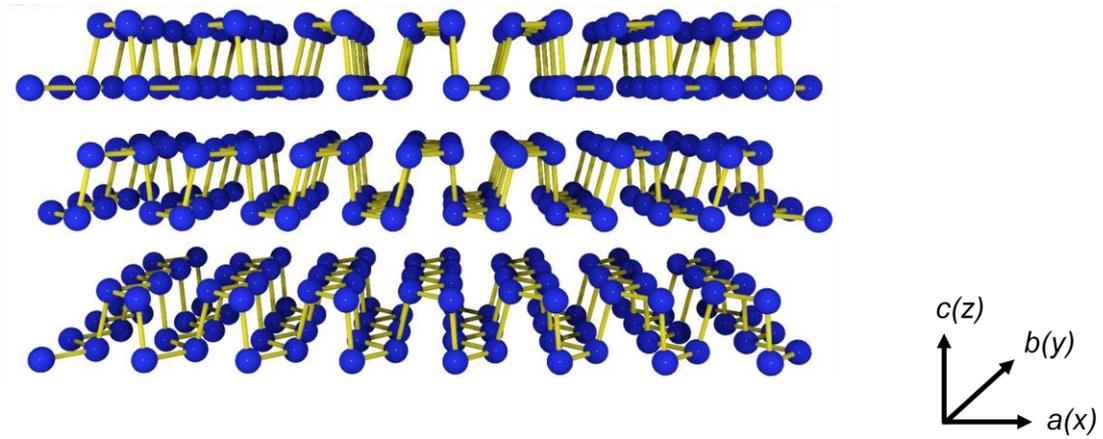

**Figure 2**

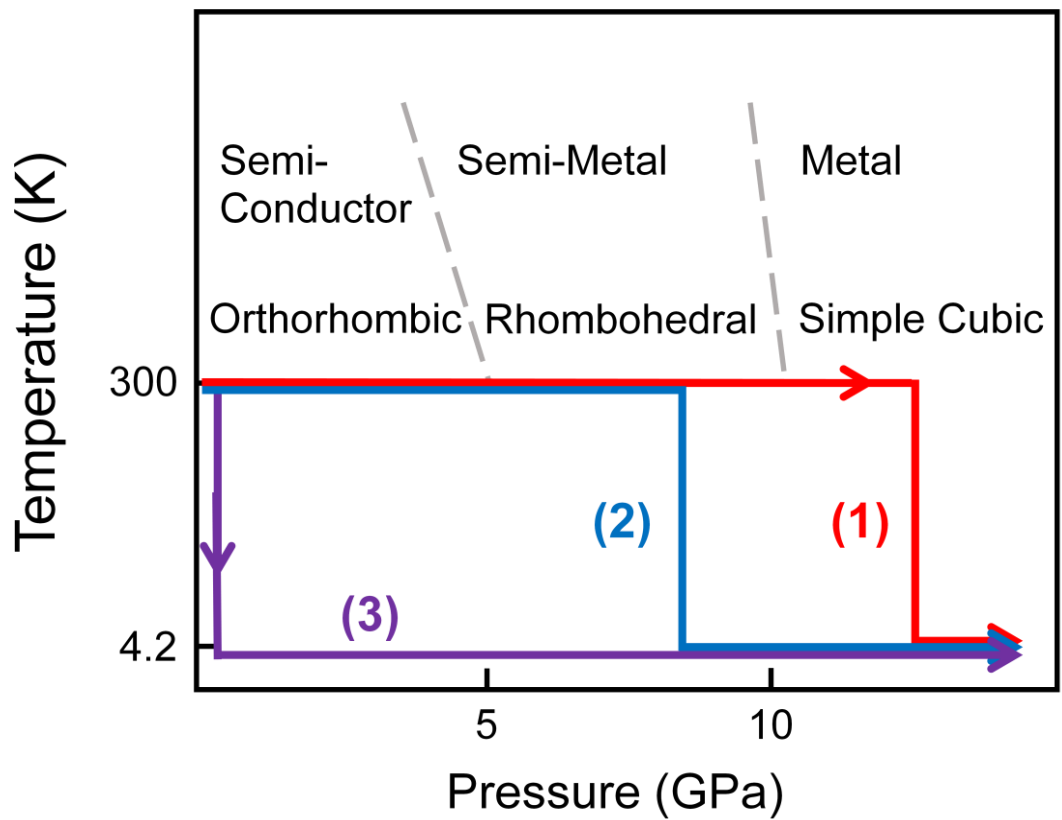

**Figure 3**

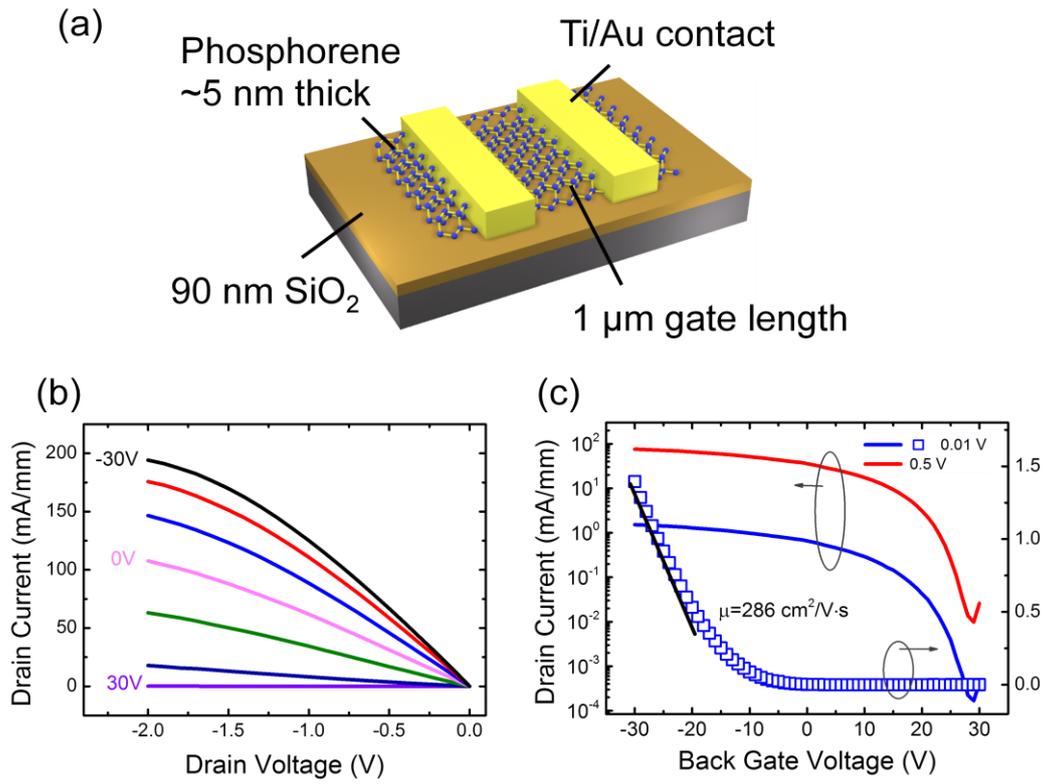

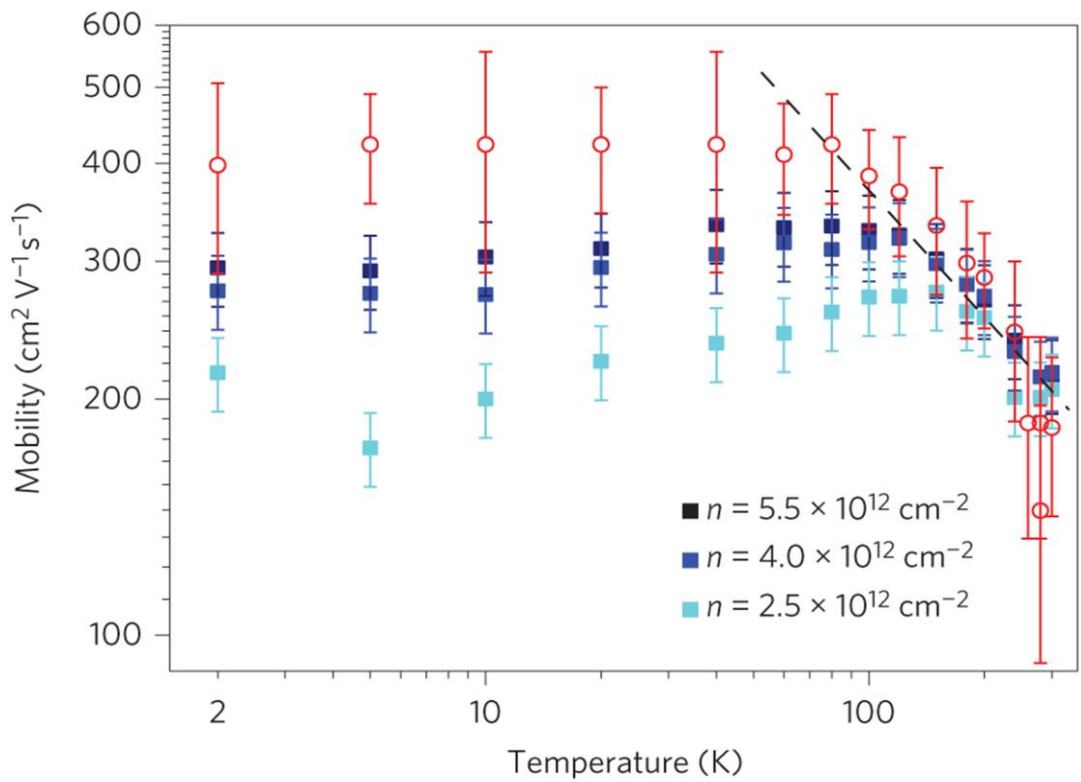

Figure 4

**Figure 5**

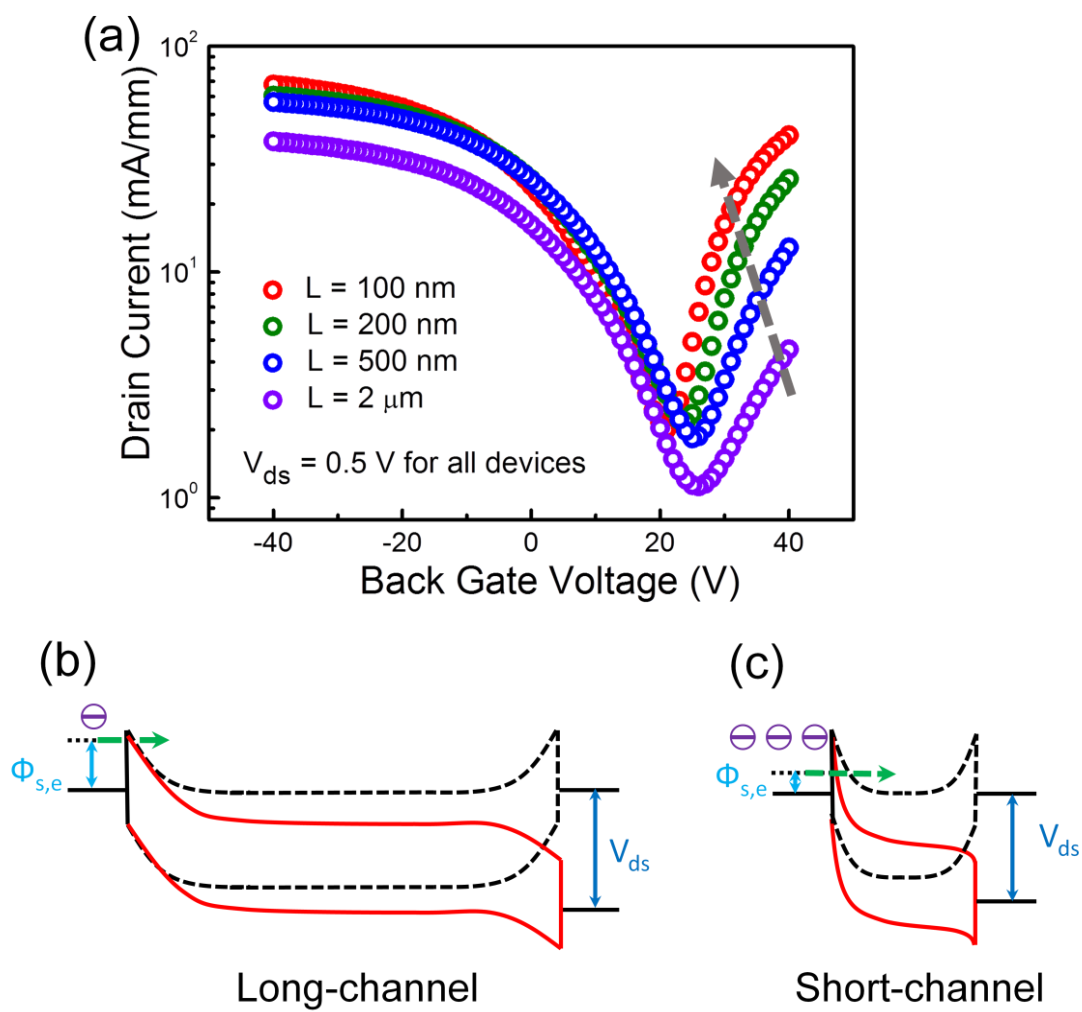

**Figure 6**

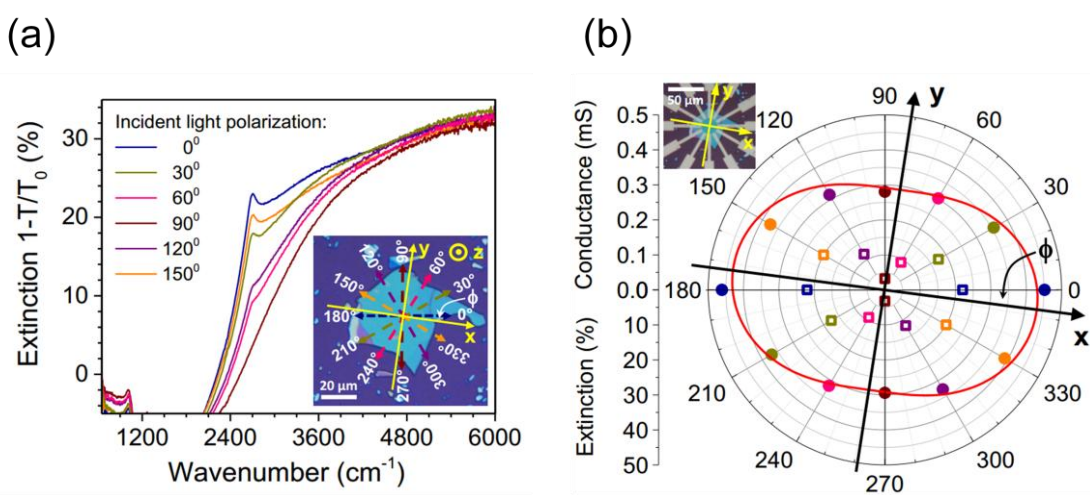

**Figure 7**

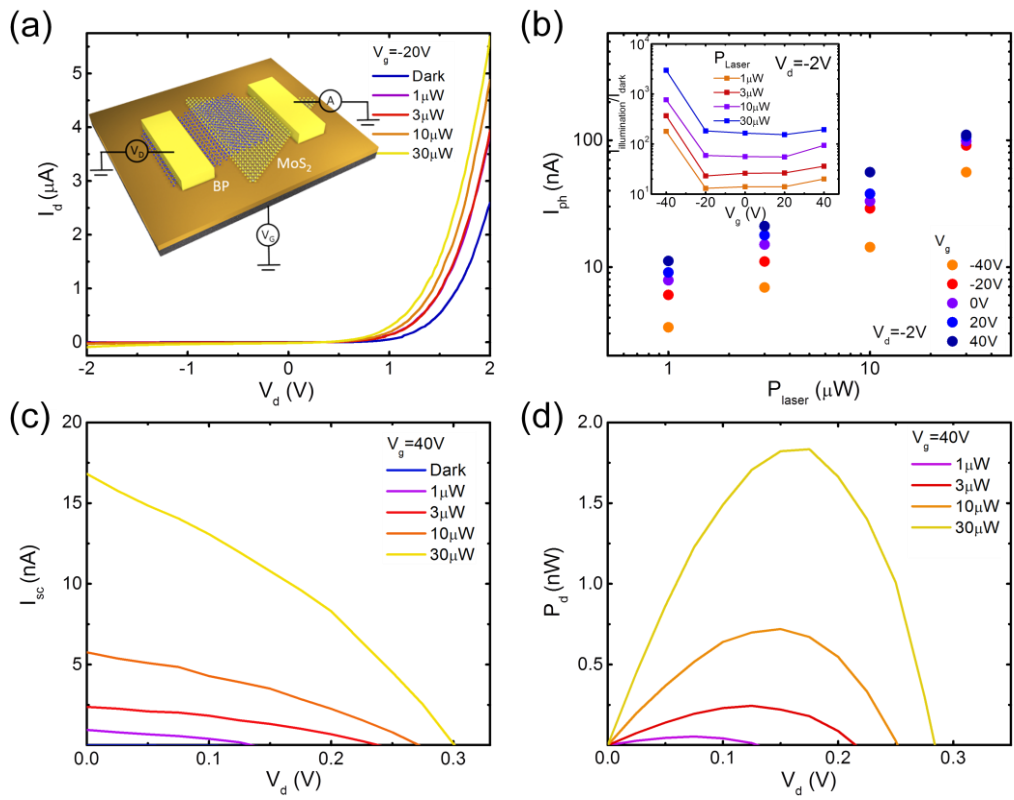

# Figure Captions

**Figure 1:** Crystal structure of few-layer black phosphorus.

**Figure 2:** Schematic phase diagram of black phosphorus. Measurement on the pressure dependence of the superconducting transition temperature are shown in three different paths. Adapted with permission from ref. 46. Copyright 1986 Springer.

**Figure 3:** (a) Schematic image of a back gate black phosphorus field effect transistor. (b) Output and (c) Transfer characteristics of black phosphorus field effect transistor. Adapted with permission from ref [11]. Copyright 2014 American Chemical Society.

**Figure 4:** Field-effect mobility (red open circles) and Hall mobility (filled squares, three different values of n) as a function of temperature on a logarithmic scale. A power-law dependence $\mu \approx T^{-0.5}$ (black dashed line) is plotted in the high-temperature region. Adapted with permission from ref [12]. Copyright 2014 Nature Publisher.

**Figure 5:** (a) Transfer characteristic black phosphorus transistors with channel length from 2 μm to 100 nm. (b)(c) Band diagram showing how Schottky barrier at metal contacts varies the transistor behavior in long and short channel transistors.

**Figure 6:** (a) Angle-resolved infrared extinction spectra when light is polarized along the six directions as shown in the inset. (b) DC conductivity and IR extinction measured along the same six directions on this black phosphorus flake and plotted in polar coordinates. Adapted with permission from ref [13]. Copyright 2014 Nature Publisher.

**Figure 7:** (a) I-V characteristics of the p-n diode under various incident laser powers. The inset shows the details in the reverse bias region. (b) The photocurrent as a function of incident laser power. Increasing the back gate voltage can increase the photocurrent. (c) $I_d$ as a function of function of $V_d$ under various laser powers. The $I_{sc}$ and $V_{oc}$ can be obtained from the intercepts of the curves on $I_d$ and $V_d$ axes. (d) Power generated by the p-n diode as a function of $V_d$ under different laser power. Adapted with permission from ref [55]. Copyright 2014 American Chemical Society.

# Biography

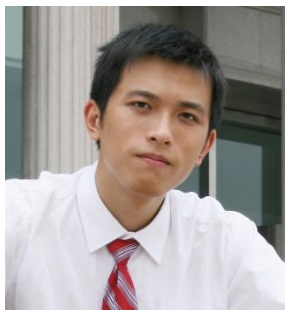
Han Liu received B.S. (2007) and M.S. (2010) in microelectronics from Fudan University, Shanghai, China. From 2010 to 2014, he was a Ph.D. student under supervision of Prof. Ye in School of Electrical and Computer Engineering of Purdue University. His research interest focused on the material and device aspects of 2D semiconducting crystals. He has been working with Intel Corporation as a device engineer since 2014.

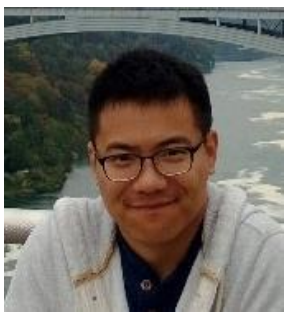
Yuchen Du received his B.S. degree in Physics from Arizona State University in 2012. He then joined the research group of Professor Peide D. Ye at Purdue University, School of Electrical and Computer Engineering. His current research interest focuses on the 2D semiconducting materials, nano-electronic devices, and magneto-transport.

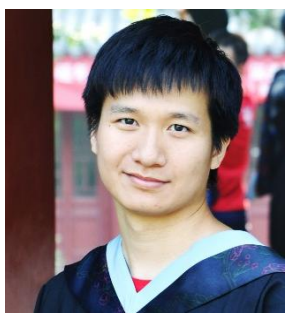
Yexin Deng received the B.S. degree from Peking University, Beijing, China, in 2013. He is currently working toward the Ph.D. degree with the School of Electrical and Computer Engineering, Purdue University, West Lafayette, Indiana, USA. His past research activities include design and simulation of 2D and 3D RRAM array architectures. He is currently working on 2D semiconductor electronic and optoelectronic devices.

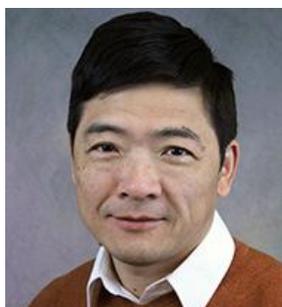
Dr. Peide Ye is a Professor of Electrical and Computer Engineering and University Faculty Scholar at Purdue University in USA. He received the B.S. from Fudan University, Shanghai, China, in 1988 and Ph.D. from Max-Planck-Institute of Solid State Research, Stuttgart, Germany, in 1996. Before joining Purdue faculty in 2005, he worked for NTT, NHMFL/Princeton University, and Bell Labs/Agere Systems. His current research is focused on ALD high-k integration on novel channel materials include III-V, Ge, CNTs and graphene, complex oxides, topological insulators, and 2D crystals. He authored and co-authored more than 300 peer reviewed articles and conference presentations. He is a Fellow of IEEE.